\documentclass[aps, prb, 10pt, twocolumn, superscriptaddress]{revtex4-1}
\usepackage[utf8]{inputenc}
\usepackage{epsfig}
\usepackage{graphicx}
\usepackage{float}
\usepackage{dcolumn}
\usepackage{bm}
\usepackage{amsmath,bbm}
\def\bra#1{\langle{#1}|}
\def\ket#1{|{#1}\rangle}

\begin{document}

\title{Magnetic Field Evolution of Spin Blockade in Ge/Si Nanowire Double Quantum Dots}

\author{A. Zarassi}\thanks{These authors contributed equally to this work.}
\affiliation{Department of Physics and Astronomy, University of Pittsburgh, Pittsburgh, PA 15260, USA} 

\author{Z. Su}\thanks{These authors contributed equally to this work.}
\affiliation{Department of Physics and Astronomy, University of Pittsburgh, Pittsburgh, PA 15260, USA} 

\author{J. Danon}
\affiliation{Department of Physics, NTNU, Norwegian University of Science and Technology, 7491 Trondheim, Norway}

\author{J. Schwenderling}
\affiliation{Department of Physics and Astronomy, University of Pittsburgh, Pittsburgh, PA 15260, USA} 
\affiliation{RWTH Aachen University, 52062 Aachen, Germany}

\author{M. Hocevar}
\affiliation{Institut N\'eel, CNRS, F-38000 Grenoble, France}

\author{B. M. Nguyen}
\affiliation{Center for Integrated Nanotechnologies, Los Alamos National Laboratory, Los Alamos, NM 87545, USA}

\author{J. Yoo}
\affiliation{Center for Integrated Nanotechnologies, Los Alamos National Laboratory, Los Alamos, NM 87545, USA}

\author{S. A. Dayeh}
\affiliation{Department of Electrical and Computer Engineering, University of California, San Diego, La Jolla,
CA 92037, USA}
\affiliation{Graduate Program of Materials Science and Engineering, University of California, San Diego, La Jolla,
CA 92037, USA}
\affiliation{Department of NanoEngineering, University of California, San Diego, La Jolla, CA 92037, USA}

\author{S. M. Frolov}\thanks{To whom correspondence should be addressed. Email: frolovsm@pitt.edu}
\affiliation{Department of Physics and Astronomy, University of Pittsburgh, Pittsburgh, PA 15260, USA} 

\date{\today}

\begin{abstract}
We perform transport measurements on double quantum dots defined in Ge/Si core/shell nanowires and focus on Pauli spin blockade in the regime where tens of holes occupy each dot. We identify spin blockade through the magnetic field dependence of leakage current. We find both a dip and a peak in the leakage current at zero field. We analyze this behavior in terms of the quantum dot parameters such as coupling to the leads, interdot tunnel coupling as well as spin-orbit interaction. We find a lower bound for spin-orbit interaction with $l_{\rm so}=500$~nm. We also extract large and anisotropic effective Land$\rm \acute{e}$ $g$-factors, with larger $g$-factors in the direction perpendicular to the nanowire axis in agreement with previous studies and experiments but with larger values reported here. 
\end{abstract}

\maketitle

Studies of spin blockade in quantum dots are largely motivated by the proposals to build a spin-based quantum computer~\cite{DiVincenzo&Loss}, as spin blockade can be used for qubit initialization and readout~\cite{hanson, petta}.
At the same time, spin blockade and its lifting mechanisms offer a direct insight into spin relaxation and dephasing processes in semiconductors and provide deeper understanding of interactions between spin localized in a quantum dot and its environment, be it the lattice and its vibrations or nuclear spins, spin-orbit interaction, or coupling to spins in nearby dots or in the lead reservoirs~\cite{koppens:science2005, johnson:nature2005, pfund:prl, pfund:prb2007, PhysRevB.81.201305}.

Holes in Ge/Si nanowires offer a relatively unexplored platform for such studies~\cite{pnas}. On the one hand, hyperfine interaction is expected to be greatly reduced owing to the low abundance of nonzero nuclear spin isotopes in the group IV materials~\cite{chargesensor2007}. Moreover, holes weakly couple to nuclear spins due to their p-wave Bloch wave symmetry, thus they are expected to come with longer spin relaxation times~\cite{fischer:prb2008}. Heavy/light hole degeneracy may also influence the spin blockade regime~\cite{carbon2012}. On the other hand, spin-orbit interaction is predicted~\cite{kloeffel} and suggested by experiments~\cite{soc2010, soc2012, dephasing:nanolett2014, antilocalization:prl2014} to be strong in Ge/Si core/shell nanowires. This offers a path to electrical spin manipulation ~\cite{nowack:science2007, inas:nature2010}, as well as to realizing Majorana fermions~\cite{lutchyn2010, oreg2010, mourik2012,gesi_majorana}.

In this work we perform transport measurements on electrostatically defined double quantum dots~\cite{hanson} made in Ge/Si core/shell nanowires, and detect Pauli spin blockade at several charge degeneracy points. We expand and adapt a previously developed rate equation model to analyze the magnetic-field evolution of the leakage current~\cite{Danon2009}. We also observe large and anisotropic $g$-factors in these dots, which supports recent theoretical predictions~\cite{maier:prb2013} and experimental observations~\cite{gfactor2008, gfactor2016}.

\begin{figure}[b]
\centering \includegraphics[width=8cm]{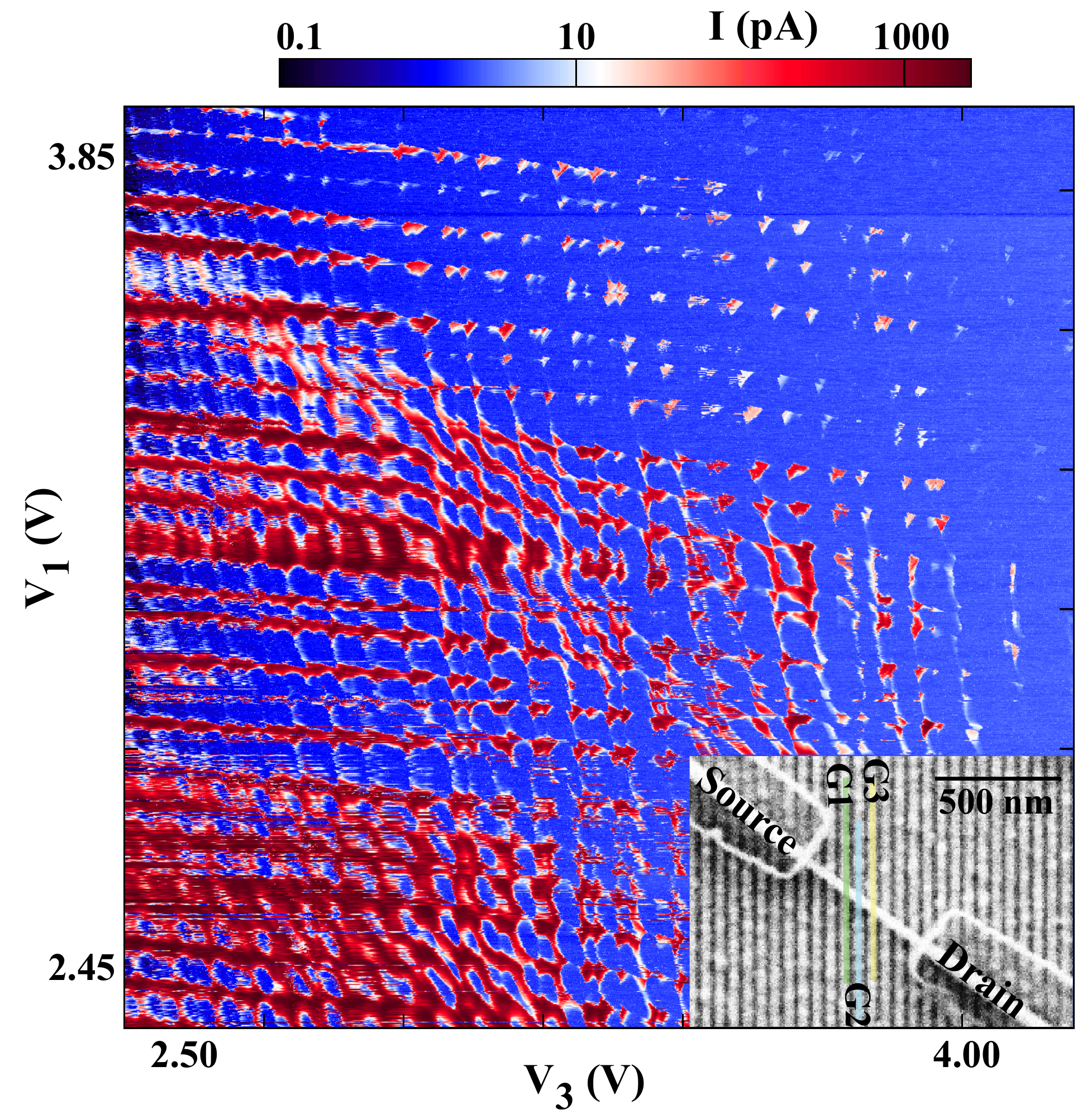}
\caption{Measured leakage current across the dots from one reservoir to the other while scanning G1 versus G3 at a fixed G2. The measurement is taken with a source-drain bias of 4~mV and at zero magnetic field. The inset shows a scanning electron micrograph of a nanowire with Al/NbTiN lithographic contacts placed on top of gold electrodes labeled G1 to G3. The other gates are not used here.
\label{fig1}}
\end{figure}

The devices are fabricated on n-doped Si substrates covered with 500~nm of thermal SiO$_2$ and patterned with local gate arrays of  Ti/Au stripes with center to center distance of 60~nm. The gates are covered by a 10~nm layer of HfO$_2$ dielectric. Using a micromanipulator~\cite{manipulation2011} the nanowires with a typical length of 4~$\mu$m and diameter of 30~nm are placed on top of these gates as shown in the inset of Fig.~\ref{fig1}. After wet etching with buffered hydrofluoric acid, we sputter 15~nm of Al followed by 42~nm of NbTiN on lithographically defined source and drain electrodes to make ohmic contacts along with the contacts to the gates. We note that despite Al and NbTiN are both superconductors the contact between the leads and the nanowire is not highly transparent, therefore the induced superconductivity is weak~\cite{arXiv}. Moreover, the proximity effect is further suppressed by the high barrier between the dots and the leads.
The measurements are performed in a dilution refrigerator at a base temperature of 30~mK.

The double quantum dot is defined by applying positive voltages to three adjacent gates: G1 and G3 are used to set the outer barriers, G2 is used to control the coupling between the dots. All three gates influence charge occupation of the dots.
The main panel of Fig.~\ref{fig1} shows the charge stability diagram of the double dot system. Many charge transitions are observed before the gate-induced energy barriers to the source and drain get too high to detect the current at the positive gate voltage extremes of the plot. This is in strong contrast with quantum dots defined using similar gates in InAs~\cite{PhysRevB.81.201305} or InSb~\cite{prl2012} nanowires, where only a few charge degeneracy points are visible between complete pinch off and the open transmission regime. Also in contrast with III-V dots, the current is too low to measure at triple points corresponding to the last few holes in these Ge/Si nanowires. Thus, in the regime studied here both dots contain tens of holes. This behavior is consistent with the large effective hole masses compared to those of electrons in III-V semiconductors.

In order to identify regimes of Pauli spin blockade we compare a charge stability diagram such as shown in Fig.~\ref{fig1} with an analogous one obtained at the opposite source-drain bias voltage, and select the charge degeneracy triangles that show a suppressed current close to their base for one of the two bias directions~\cite{ono2002} (see supplementary information). We then focus on these regimes of suppressed current and investigate the evolution of the current in the presence of an externally applied magnetic field, which provides additional evidence of spin blockade.

Although the charge transitions studied here are $(n,m) \to (n-1,m+1)$ with $n,m \sim 10$ the hole occupations of the two dots, we assume that the spin blockade we find can be effectively understood in the same way as the simplest $(1,1) \to (0,2)$ spin blockade:
Close to the base of the bias triangle, the $n$'th hole in the source dot can only enter the drain dot if it can form a singlet state with the $m$'th hole on the drain dot. Entering an $(n-1,m+1)$ state in a triplet configuration requires occupation of a higher orbital state which becomes energetically accessible only when an additionally applied interdot detuning exceeds the single-dot orbital level splitting in the drain dot.
For small detuning the system is thus expected to be blocked in one of the three triplet states, which are in principle degenerate and split in energy under the influence of a magnetic field due to the Zeeman effect.
For clarity we will refer to the $(n,m)$ states as $(1,1)$ and to the $(n-1,m+1)$ states as $(0,2)$.

\begin{figure}[t]
\centering \includegraphics[width=7cm]{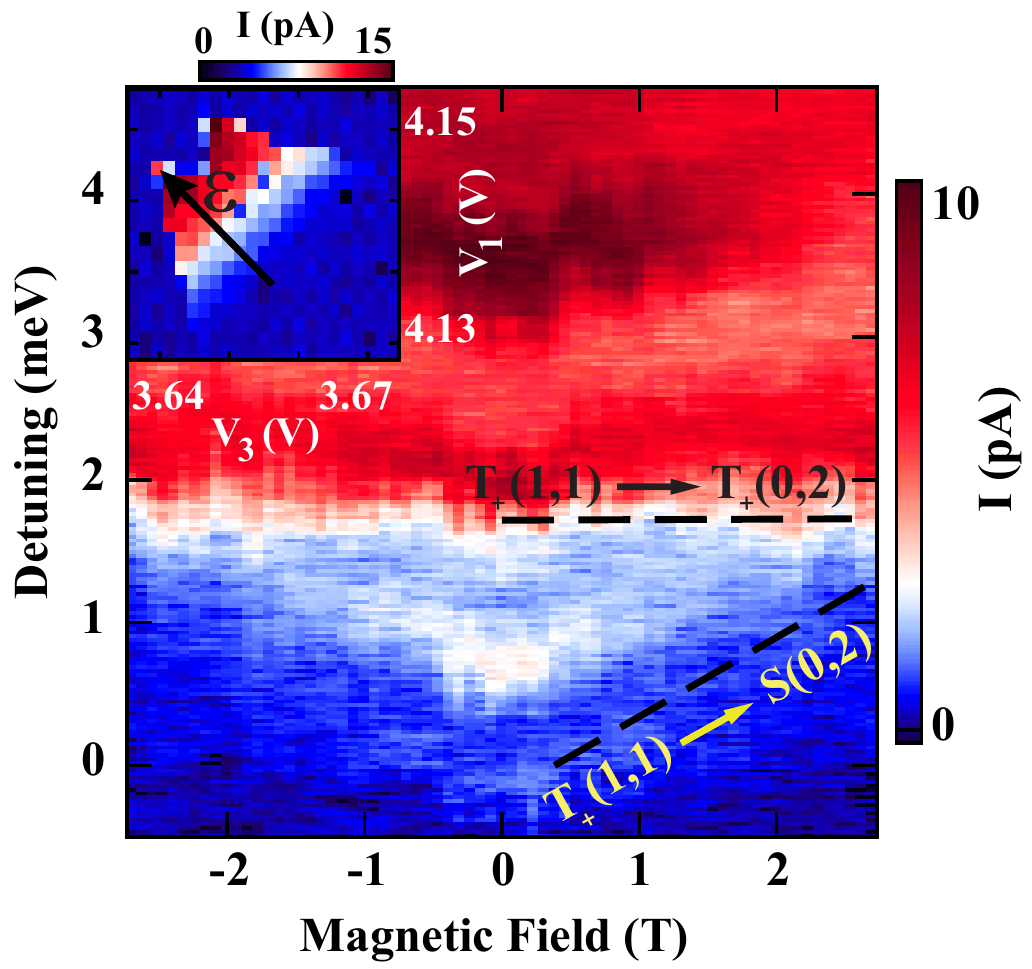}
 \caption{
Leakage current through the double quantum dot, measured as a function of detuning $\varepsilon$ and magnetic field $B$, with an applied source-drain voltage of $V_{SD}=6.5$~mV. The magnetic field is applied normal to the substrate plane. The resonances associated with the $T_+(1,1) \to T_+(0,2)$ and $T_+(1,1) \to S(0,2)$ transitions are marked with dashed lines. From the field dependence of the latter we find $g=9.2$. Inset: the bias triangle with the detuning axis indicated. \label{fig2}  }
\end{figure}

First, we use our field-dependent data to investigate the effective hole $g$-factors.
In Fig.~\ref{fig2} we show the leakage current inside one of the bias triangles as a function of magnetic field (applied perpendicular to the plane of the nanowire and gates) and the (1,1) to (0,2) energy level detuning, $\varepsilon$.
We vary $\varepsilon$ by scanning G1 and G3 perpendicular to the base of bias triangles (as indicated in the inset), while stepping the magnetic field.
The suppressed current observed for $0 < \varepsilon \lesssim 2$~meV is associated with spin blockade, and we interpret the sudden rise in current at $\varepsilon \approx 2$~meV as the $(0,2)$ triplet states becoming energetically accessible from the $(1,1)$ triplet states, thus lifting the blockade.
The associated singlet-triplet splitting of $\sim 2$~meV is representative of the several charge degeneracy points studied (see supplementary information).

At finite magnetic field, the increase in current is expected when the lowest triplet states $T_+(1,1)$ and $T_+(0,2)$ are resonant (see the upper dashed line in Fig.~\ref{fig2}).
A finite slope of this resonance as a function of $B$ would reflect a $B$-dependent energy difference between $T_+(1,1)$ and $T_+(0,2)$, indicating a difference between the effective $g$-factors on the two dots. We do not observe a significant $B$-dependence, and the effective $g$-factors on the two dots thus cannot be distinguished within the resolution of this measurement.
An upper bound for the $g$-factor difference in the bias triangle of Fig.~\ref{fig2} can be read off as $\frac{1}{2}|g_L-g_R| \lesssim 0.8$.

\begin{figure}[t]
 \centering \includegraphics[width=8.75cm]{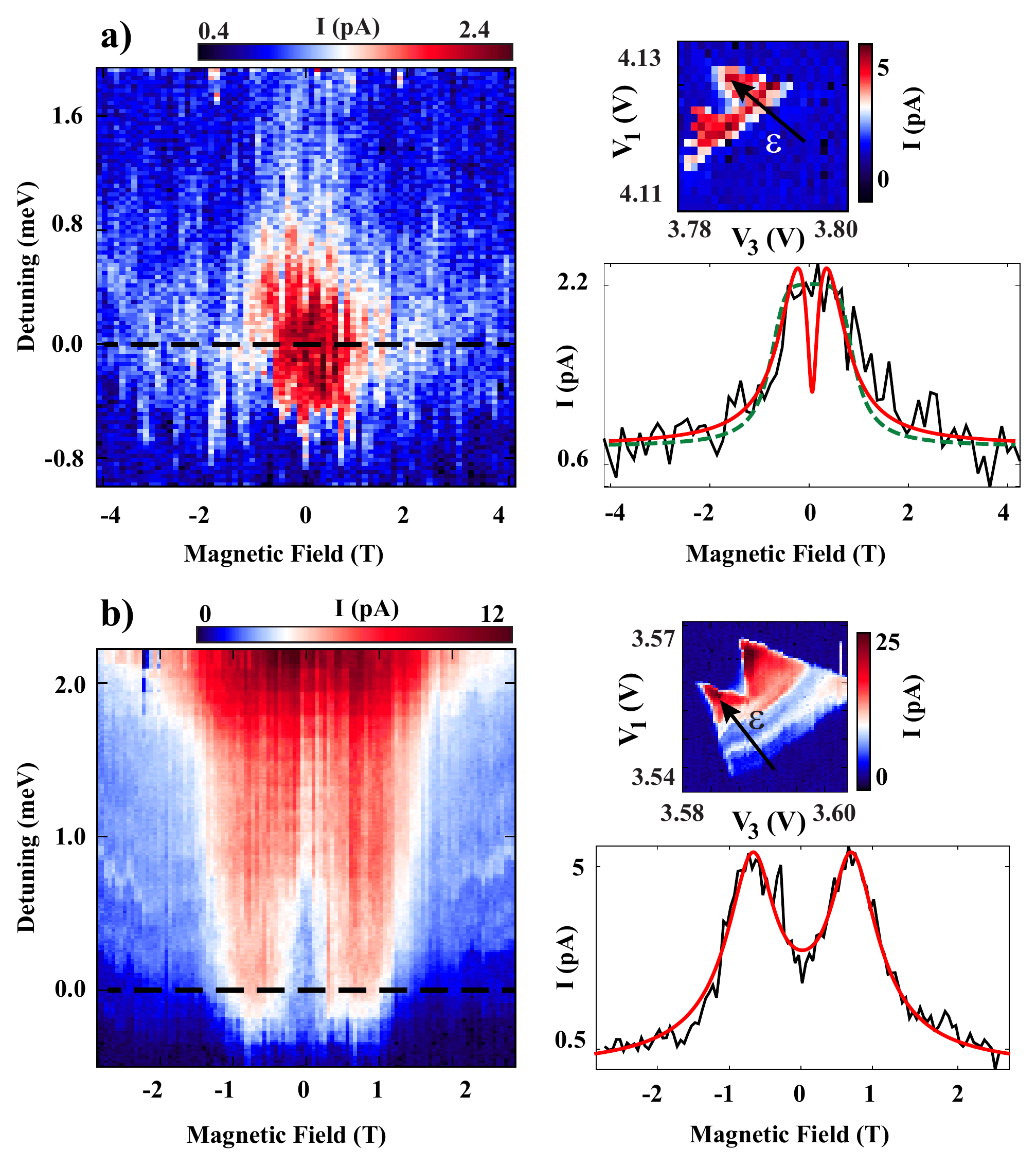}
 \caption{
 Magnetic field evolution of the leakage current in two different spin-blockaded transport configurations.
 In both cases the field is applied in the plane of the nanowire and gates, perpendicular to the gates but making an angle of $\sim 30^{\circ}$ with the wire.
 In the left panels we show the dependence of the leakage current on magnetic field and detuning, and on the right side we show the corresponding bias triangles (top) and a line cut of the data at zero detuning (bottom).
 The zero-detuning cuts include fits to the theory presented in the main text.
 (a) In this configuration, where a bias voltage $V_{SD}=6.5$~mV is applied, the leakage current has a single-peak structure both as function of detuning and magnetic field.
 The corresponding bias triangle is taken at $B=5$~T.
 In the figure we plot two different theory curves on top of the data, both with $\xi = 0.03$, $g=5$, and an added constant current of $0.8~$pA to account for the background signal observed in the data.
 We further used $\Gamma = 300$~MHz, $t = 55~\mu$eV, $\gamma = 0.007$, and $\alpha = 0.4$ (solid red curve) and $\Gamma = 25$~MHz, $t = 170~\mu$eV, $\gamma = 0.7$, and $\alpha = 0.4$ (dashed green curve).
 (b) Leakage current in a different bias triangle, with $V_{SD}=4$~mV. The corresponding bias triangle is taken at $B=0$~T, where the suppressed current at the base of the triangle is evident. 
 Here the current shows a double-peak structure, which can also be clearly seen in the zero-detuning cut.
 The theory curve (red solid line) uses $\xi = 0.03$, $g=4.4$, $\Gamma = 256$~MHz, $t = 165~\mu$eV, $\gamma = 0.061$, and $\alpha = 0.37$.
 \label{fig3}
 }
\end{figure}

A smaller rise in the leakage current at lower detuning, marked with the tilted dashed line in Fig.~\ref{fig2}, is assigned to a resonance between the lowest $(1,1)$ state $T_+$ and the singlet $S(0,2)$ state:
Below this resonance (for smaller $\varepsilon$), $S(0,2)$ is energetically not accessible from the ground state $T_+(1,1)$ and the system is in Coulomb blockade.
Since the energy of $S(0,2)$ is not expected to depend on the magnetic field, the $B$-dependence of this resonance reflects the $B$-dependence of the energy of $T_+(1,1)$.
Hence, we can use the slope of this resonance to read off the average effective hole $g$-factor of the two dots. Note that several copy resonances follow the $T_+(1,1) \to S(0,2)$ transition in field, these resonances are not accounted for in the simple spin-blockade picture used here. For Fig.~\ref{fig2}, we obtain $g=9.2$, while other triple points (Fig.~\ref{fig3}) yield lower values such as 4.5 and 5. While full $g$-tensor measurements were not performed, we find lower $g$-factors for fields deviating from normal to the substrate, in agreement with other studies (see supplementary information)~\cite{gfactor2008, gfactor2016}. Overall, the $g$-factors measured here are larger than previously reported for Ge/Si nanowires~\cite{soc2012, gfactor2008, gfactor2016}. One possible reason for this is that wires of larger diameters were used here. In small-diameter wires, effective $g$-factors can be reduced towards the free electron $g$-factor due to orbital quenching~\cite{pryorprb2012}.

We now turn to a more detailed investigation of the spin blockade.
In Fig.~\ref{fig3}a,b (left panels) we plot the measured leakage current in the spin-blockade regime of two representative bias triangles which show a qualitatively different field-dependent behavior.
The current in Fig.~\ref{fig3}a shows a single peak centered at zero field.
In Fig.~\ref{fig3}b, we clearly observe a different behavior of the leakage current: a double peak structure with a dip at zero magnetic field.
We note that beyond the difference in charge numbers, we cannot independently quantify differences in other double dot parameters across the two regimes of Fig.~\ref{fig3}. We speculate that the interdot tunnel coupling as well as the couplings to the leads are not the same in the two regimes, as the three adjacent gates have relatively strong cross coupling to each other and even slight changes on them can reshape the double quantum dot configuration.

A zero-field dip in the leakage current is known to occur in double dots hosted in materials with strong spin-orbit interaction~\cite{pfund:prl,PhysRevB.81.201305,Yamahata2012,Li2015,Bohuslavskyi2016}.
The dip is usually explained in terms of a competition between different types of spin-mixing processes:
The combination of spin-orbit interaction and Zeeman splitting due to the applied field enables transitions between triplet and singlet configurations.
This mechanism becomes more efficient at higher magnetic field and thus it produces a dip in the leakage current around zero field~\cite{Danon2009}.
Other processes that mix spin states, such as the hyperfine interaction between the electrons or holes and the nuclear spins in the host material~\cite{jouravlev:prl} or spin-flip cotunneling processes with the leads~\cite{Coish2011}, can be independent of the magnetic field or even become less efficient with increasing $B$.
If one of such processes provides the dominant spin-mixing mechanism, then there will appear no dip in the current around zero field.
Since the spin-orbit-mediated mechanism scales with the interdot tunnel coupling, one can  expect to observe a transition from having a zero-field dip to no zero-field dip when changing the tuning of the double dot.

Ignoring the potentially more complicated nature of spin blockade in the valence band, we assume that in the present case we can describe the leakage current with a model based on the following ingredients:
(i) $S(1,1)$, has the same singlet configuration as $S(0,2)$ and is thus strongly coupled to that state, with a coupling energy $t$.
(ii) The state $S(0,2)$ decays to the drain lead with a rate $\Gamma$.
Immediately after such a transition a new hole enters the system from the source, bringing it in one of the $(1,1)$ states again.
(iii) $T_{\pm}(1,1)$ split off in energy when a magnetic field is applied.
(iv) Spin-orbit interaction results in a coherent non-spin-conserving coupling between the $(1,1)$ triplet states and $S(0,2)$.
The energy scale characterizing spin-orbit coupling $t_{\rm so}$ is proportional to $t$.
(v) There can be other spin-mixing and spin-relaxation processes causing transitions between the different $(1,1)$ states.

One issue, however, that sets our data apart is that both the dip and the peak we observe are relatively wide: they appear on a field scale of $B \sim 1$~T which is of the order of 3~K.
First of all, this rules out hyperfine interaction as the dominant spin-mixing mechanism in the single-peak data of Fig.~\ref{fig3}a. Hyperfine interaction is known to lift spin-blockade around zero field producing a peak in current, but the width of the hyperfine peak is comparable to the typical magnitude of the effective nuclear fields in the dots.
We estimate the effective nuclear fields in the present system to be less than 10 mT, which is orders of magnitude smaller than the peak width observed here~\cite{hfge1966}.
Secondly, the analytic theory of Ref.~\onlinecite{Danon2009}, which is often used to extract model parameters such as the magnitude of spin-relaxation rates and $\alpha = t_{\rm so}/t$, is valid for $t,t_{\rm so},B \ll \Gamma$ and also assumes the spin-relaxation rates to be isotropic, based on the assumption $B \ll T$. From here on we will use $\hbar = k_{\rm B} = g\mu_{\rm B} = e = 1$.
In the present case, however, we have $B \gg T$ for most fields of interest, and spin relaxation will thus mostly be directed towards the $(1,1)$ ground state instead.
Furthermore, the suppression of current at the highest fields could indicate that $B$ exceeds at these fields the effective level width of $S(0,2)$ by such an amount that the system is pushed into a Coulomb blockade in the lowest-lying $(1,1)$ triplet state.

We thus cannot straightforwardly apply the theory of Ref.~\onlinecite{Danon2009} to model the data shown in Fig.~\ref{fig3}.
Instead we present a modified version of the theory, where we use a unidirectional spin-relaxation rate and do not expand in large $\Gamma$.
We start from the five-level Hamiltonian
\begin{align}
H = \left( \begin{array}{ccccc}
0 & iB & 0 & 0 & i\alpha t \\
-iB & 0 & 0 & 0 & i\alpha t \\
0 & 0 & 0 & 0 & i\alpha t \\
0 & 0 & 0 & 0 & t \\
-i\alpha t & -i\alpha t & -i\alpha t & t & 0
\end{array}\right),\label{eq:ham}
\end{align}
written in the basis $\{ \ket{T_x}, \ket{T_y}, \ket{T_z}, \ket{S}, \ket{S_{02}} \}$, where $\ket{T_{x,y}} = i^{1/2 \mp 1/2}\{\ket{T_-} \mp \ket{T_+}\}/\sqrt 2$ and $\ket{T_z} = \ket{T_0}$ are the three $(1,1)$ triplet levels and $\ket{S}$ and $\ket{S_{02}}$ the $(1,1)$ and $(0,2)$ singlets, respectively.
The interdot detuning was set to zero and $\alpha$ parametrizes the strength of the effective spin-orbit interaction in the dots, where $\alpha \sim 1$ corresponds to the strong limit.
In principle, the three $\alpha$'s coupling of $\ket{T_{x,y,z}}$ to $\ket{S_{02}}$ can be different, constituting a vector $\boldsymbol\alpha = (\alpha_x, \alpha_y, \alpha_z)$ (see Ref.~\onlinecite{Danon2009}). The length of this vector corresponds to the strength of the spin-orbit interaction and its direction is related to the direction of the effective spin-orbit field. In a physical nanowire, the precise orientation of $\boldsymbol\alpha$ depends on many details and is hard to predict. We therefore make the simplifying assumption that all three components are of the same magnitude.
We diagonalize the Hamiltonian and use its eigenbasis to write a time-evolution equation for the density matrix~\cite{Danon2009},
\begin{align}
\frac{d\hat\rho}{dt} = -i[H^{\rm diag},\hat\rho ] + \boldsymbol \Gamma \hat\rho + \boldsymbol \Gamma_{\rm rel} \hat\rho.
\label{eq:drdt}
\end{align}
The operator $\boldsymbol\Gamma$ describes (i) decay of all states $\ket{n}$ (with $n=0\dots 4$) to the drain lead with the rates $\Gamma |\langle {n|S_{02}} \rangle|^2$ and (ii) immediate reload into one of the eigenstates with the probabilities $\{ 1 - |\langle {n|S_{02}} \rangle|^2 \}/4$.
For the relaxation operator $\boldsymbol\Gamma_{\rm rel}$ we take a simple form:
We assume that all four excited states relax with the same rate $\Gamma_{\rm rel}$ to the ground state.
At $B=0$ this ground state is an equal superposition of $\ket{S_{02}}$ and the optimally coupled $(1,1)$ state $\ket{m} = \{\ket{S} - i\alpha \mathbbm{1}\cdot\ket{\vec T}\}/\sqrt{1+3\alpha^2}$, and for $B\to \infty$ it develops into a pure $\ket{T_+}$-state.

We first discuss this model on a qualitative level, and investigate how it differs from the model of Ref.~\onlinecite{Danon2009}.
For small fields, $B \ll \Gamma$, the unidirectionality of the relaxation processes only yields different numerical factors in some of the results.
At $B=0$ we have three blocked states at zero energy that can relax to the hybridized $(1,1)$--$(0,2)$ ground state which quickly decays to the drain lead; this results on average in four holes being transported through the system in a time $3\Gamma_{\rm rel}^{-1}$, thus yielding a leakage current of $I(0)=\frac{4}{3}\Gamma_{\rm rel}$.
Adding a finite magnetic field induces a coupling of $\sim \alpha B$ between two of the blocked states and $\ket{m}$, which provides an alternative escape route and leads to an increase of the current.

This increase becomes significant only when the rate of this escape $\sim (\alpha B)^2\Gamma/t^2$ becomes comparable to $\Gamma_{\rm rel}$, which happens at $B \sim (t/\alpha)\sqrt{\Gamma_{\rm rel}/\Gamma}$.
For larger fields the current tends to its maximum value $I_{\rm max} = 4\Gamma_{\rm rel}$, reached when only one truly blocked state is left and on average four holes are transported in a time $\Gamma_{\rm rel}^{-1}$.
We see that this picture predicts a zero-field dip in the current of width $B_{\rm dip} \sim (t/\alpha)\sqrt{\Gamma_{\rm rel}/\Gamma}$ and a maximal suppression of the current, by a factor 3, at $B=0$.
This is, apart from numerical factors, the same result as found in Ref.~\onlinecite{Danon2009}.

Qualitative differences appear when we investigate what happens at even higher fields.
Since $\Gamma$ is finite in the present model and relaxation is unidirectional, we can enter a situation of Coulomb blockade in the $(1,1)$ ground state $\ket{T_+}$.
When we increase $B$, the current will thus eventually be suppressed to zero, producing in general a double-peak structure in $I(B)$.
A na{\"i}ve guess for the field scale where this suppression sets in would be $\sim \Gamma$:
The level width of $\ket{S_{02}}$ is set by $\Gamma$, and for $B\gtrsim \Gamma$ the escape rate from $\ket{T_+}$ drops gradually to zero.
However, the actual field scale of current decay is rather set by the competition of this escape rate with $\Gamma_{\rm rel}$:
Only when the $B$-induced suppression becomes so strong that escape from $\ket{T_+}$ is the main bottleneck for the leakage current, the decrease in current becomes significant.
We thus compare this escape rate $\sim (\alpha t)^2 \Gamma / B^2$ with $\Gamma_{\rm rel}$ and find an estimate for the width of the overall double-peak structure $B_c \sim \alpha t \sqrt{\Gamma/\Gamma_{\rm rel}}$.

We can also understand how our model could result in an apparent single-peak $I(B)$.
Indeed, $B_{\rm dip}$ and $B_c$ show a different dependence on the model parameters, and their ratio $B_{\rm dip}/B_c \sim \Gamma_{\rm rel}/\alpha^2 \Gamma$ (which determines the relative visibility of the zero-field dip) could be large or small, depending on the detailed tuning of all parameters.
For $B_{\rm dip}/B_c \ll 1$ one could be in the situation where the central dip around zero field is too narrow to be observed. 

We will now support these arguments with a more quantitative investigation of the model.
We can solve Eq.~\ref{eq:drdt} in steady state, $d\hat\rho / dt = 0$, and find the current from the resulting equilibrium occupation probabilities $p_n = \hat\rho_{nn}$ as $I = \sum_n p_n\Gamma |\langle {n|S_{02}} \rangle|^2$, yielding
\begin{align}
I(B) = \Gamma_{\rm rel}
\frac{[w-B^2+\tau^2][w(1+4\gamma) + B^2-\tau^2]}{6 \gamma  w^2 + 2 B^2 \alpha^2t^2},
\label{eq:curr}
\end{align}
where we use the notation $w = \sqrt{(B^2 - \tau^2)^2 + 8B^2\alpha^2t^2}$, the small parameter $\gamma = \Gamma_{\rm rel} / \Gamma$, and $\tau = t \sqrt{1 + 3\alpha^2}$ (which is the total tunnel coupling energy).
To obtain Eq.~\ref{eq:curr} we assumed $\gamma \ll 1$, which we will also do below.

The current given by Eq.~\ref{eq:curr} indeed shows in general a double-peak structure.
At zero field we find $I(0)=\frac{4}{3}\Gamma_{\rm rel}$, and the current has two maxima at $B = \pm \tau$ where $I = 4\Gamma_{\rm rel}$.
The half-width of the resulting zero-field dip follows as $B_{\rm dip} = t(\sqrt{\beta^2+2} - \beta)/\sqrt 2$, where $\beta = \alpha/\sqrt{6\gamma}$.
In the limit of large $\beta$ (small $\sqrt \gamma / \alpha$) we find $B_{\rm dip} \approx t\sqrt{3\gamma}/\alpha$.
At high fields, the current drops to zero, and from Eq.~\ref{eq:curr} we find the half-width-half-maximum of the full double-peak structure to be $B_{\rm c} = t(\sqrt{\beta^2+2} + \beta)/\sqrt 2$ which reduces to $B_c \approx \alpha t / \sqrt{3\gamma}$ for large $\beta$.
We see that in the limit of small $\gamma$ these results agree with the conclusions of our qualitative discussion above.

\begin{figure}[t]
 \centering \includegraphics[scale=1]{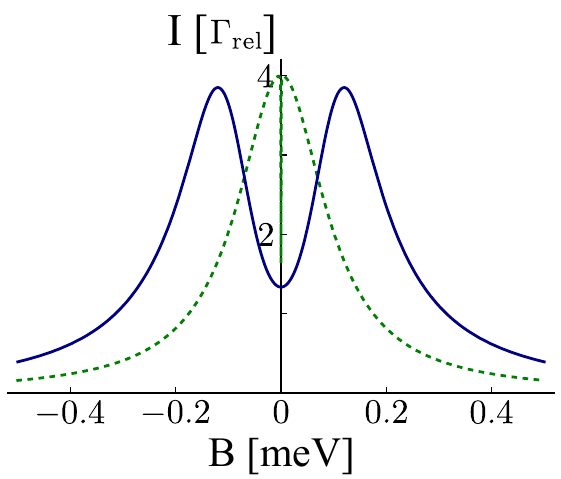}
 \caption{
 The current resulting from Eq.~\ref{eq:curr} for two different sets of parameters:
 $t = 120~\mu$eV, $\alpha = 0.1$, and $\gamma = 2\times 10^{-3}$ (solid blue curve) and $t = 3.5~\mu$eV, $\alpha = 0.5$, and $\gamma = 10^{-4}$ (dashed green curve).
 \label{fig4}
 }
\end{figure}

In Fig.~\ref{fig4} we plot $I(B)$ for two different sets of parameters, illustrating how the model can produce curves that appear to have double-peak as well as single-peak structures. The solid curve shows a clear double-peak structure, which is indeed expected since the ``visibility parameter'' $B_{\rm dip}/B_c \approx 0.30$ predicts a clearly distinguishable zero-field dip. In contrast, for the dashed curve $B_{\rm dip}/B_c \approx 0.001$.
In this case, the current still has a dip around zero field; its width, however, is $\sim 1000$ times smaller than the overall width of the structure and therefore invisible in the plot.
Depending on all other parameters, this situation could thus correspond to an experiment where the leakage current appears to have a single-peak structure.

In order to connect our model to experimental data in Fig.~\ref{fig3} and facilitate fitting of the model parameters (see below), we include the likely scenario that $g$-factors in the two dots are significantly different.
The effective $g$-factor for a localized hole depends on many microscopic characteristics, among which the details of the confining potential\cite{kloeffel}, and is thus expected to differ from dot to dot.
Based on the data shown in Fig.~\ref{fig2} we estimated the difference between the $g$-factors to be smaller than $\sim$~10\%, but a difference of 2--5\% is highly probable~\cite{Bohuslavskyi2016, voisin2016}.
The effect of having different $g$-factors on the left and right dots ($g_L$ and $g_R$) is a coherent mixing of $\ket{T_z}$ and $\ket{S}$.
As a result, the single blocked state left at finite field $\{\ket{T_z} + i\alpha \ket{S}\}/\sqrt{1+\alpha^2}$ couples to the decaying state $\{\ket{S} - i\alpha \ket{T_z}\}/\sqrt{1+\alpha^2}$, thus lifting the blockade.
The rate of this decay of the last blocked state is $\Gamma_{\xi} \sim (\xi B)^2\Gamma / t^2$, where $\xi = \frac{1}{2}(g_L - g_R)/(g_L+g_R)$. This decay competes with $\Gamma_{\rm rel}$ for being the bottleneck for the leakage current:
If $\Gamma_\xi \gtrsim \Gamma_{\rm rel}$ then the overall scale of the current will be set by $\Gamma_\xi$.
Independent of the relative importance of $\Gamma_\xi$ and $\Gamma_{\rm rel}$, this sets an upper bound on $\Gamma_\xi$ and thus on $\Gamma$ (several pA for the data shown in Fig.~\ref{fig3}).

To include the effect of a finite $g$-factor gradient into our model, we add a term $H_\xi = \xi B \{\ket{T_z}\bra{S} + \ket{S}\bra{T_z}\}$ to the Hamiltonian (\ref{eq:ham}).
We can again solve Eq.~\ref{eq:drdt} in steady state $d\hat\rho/dt = 0$ and arrive at an analytic expression for the current $I(B)$ which we can fit to the data (at this point we do not assume $\gamma \ll 1$).
Fixing $\xi = 0.03$, we can obtain reasonable fits to the double-peak data of Fig.~\ref{fig3}b (See the supplementary material for an explicit expression for $I(B)$ including a finite $\xi$). 
Based on these results, we conclude that spin-orbit parameter $\alpha$ is in the range $\sim 0.1$--$0.4$. The single-peak data of Fig.~\ref{fig3}a are harder to fit due to lack of features, thus we cannot reasonably narrow down all the fit parameters. However, theory curves with $\alpha$ in the same range as for the double-peak regime can show reasonable agreement, see Fig.~\ref{fig3}b.

To conclude, assuming linear Rashba spin-orbit interaction as the dominant relaxation term~\cite{kloeffel} in these gate-defined double quantum dots with $\alpha = 0.1$--$0.4$, 
and a dot-to-dot distance of order 50 nm, we find a spin-orbit length of $l_{\rm so} = 100$--$500$~nm.
While this corresponds to a substantial spin-orbit interaction, it does not greatly exceed that measured in InAs or InSb nanowires, contrary to some expectations~\cite{kloeffel}. One possibility for this could be that $\alpha$ is not maximal for the field orientation at which data is obtained here as a consequence of spin-orbit anisotropy~\cite{prl2012}, although magnetic field was not oriented in the direction expected for the spin-orbit field. Another factor for low-than-expected spin-orbit interaction is the low strain between the thin Si shell and relatively thick Ge core. Thus, it is conceivable that spin-orbit interaction can be enhanced by tailoring the nanowire morphology. A more detailed insight into spin-orbit coupling and other double dot parameters could be obtained from electric dipole spin resonance.

The Ge/Si nanowire growth was performed at the Center for Integrated Nanotechnologies (CINT), U.S. Department of Energy, Office of Basic Energy Sciences User Facility at Los Alamos National Laboratory (Contract DE-AC52-06NA25396) and Sandia National Laboratories (Contract DE-AC04-94AL85000). We thank T. Baron, R. Chen, S. De Franceschi, P. Gentile, D. Kotekar-Patil, E. Lee and P. Torresani for technical help and useful discussions. S.A.D. acknowledges NSF support under DMR-1503595 and ECCS-1351980. S.M.F. acknowledges NSF DMR-125296, ONR N00014-16-1-2270 and Nanoscience Foundation, Grenoble.

\bibliographystyle{unsrt}
\bibliography{Ref_main_text.bib}

\pagebreak
\widetext
\clearpage

\begin{center}
\textbf{\large Supplementary Information: Magnetic Field Evolution of Spin Blockade in Ge/Si Nanowire Double Quantum Dots}
\end{center}
\setcounter{figure}{0}
\setcounter{equation}{0}
\setcounter{page}{1}
\renewcommand{\thefigure}{S\arabic{figure}}
\renewcommand{\theequation}{S\arabic{equation}}

\tableofcontents
\section{Charge stability diagrams}

We have measured three different nanowire devices, A, B, and C, in three different dilution refrigerators all at base temperatures below 30~mK. All three devices have the same fabrication recipe as mentioned in the main text. In this supplementary information we show more data on Device A, the device in the main text, as well as data from the other two devices.

\begin{figure}
\includegraphics[width=12cm]{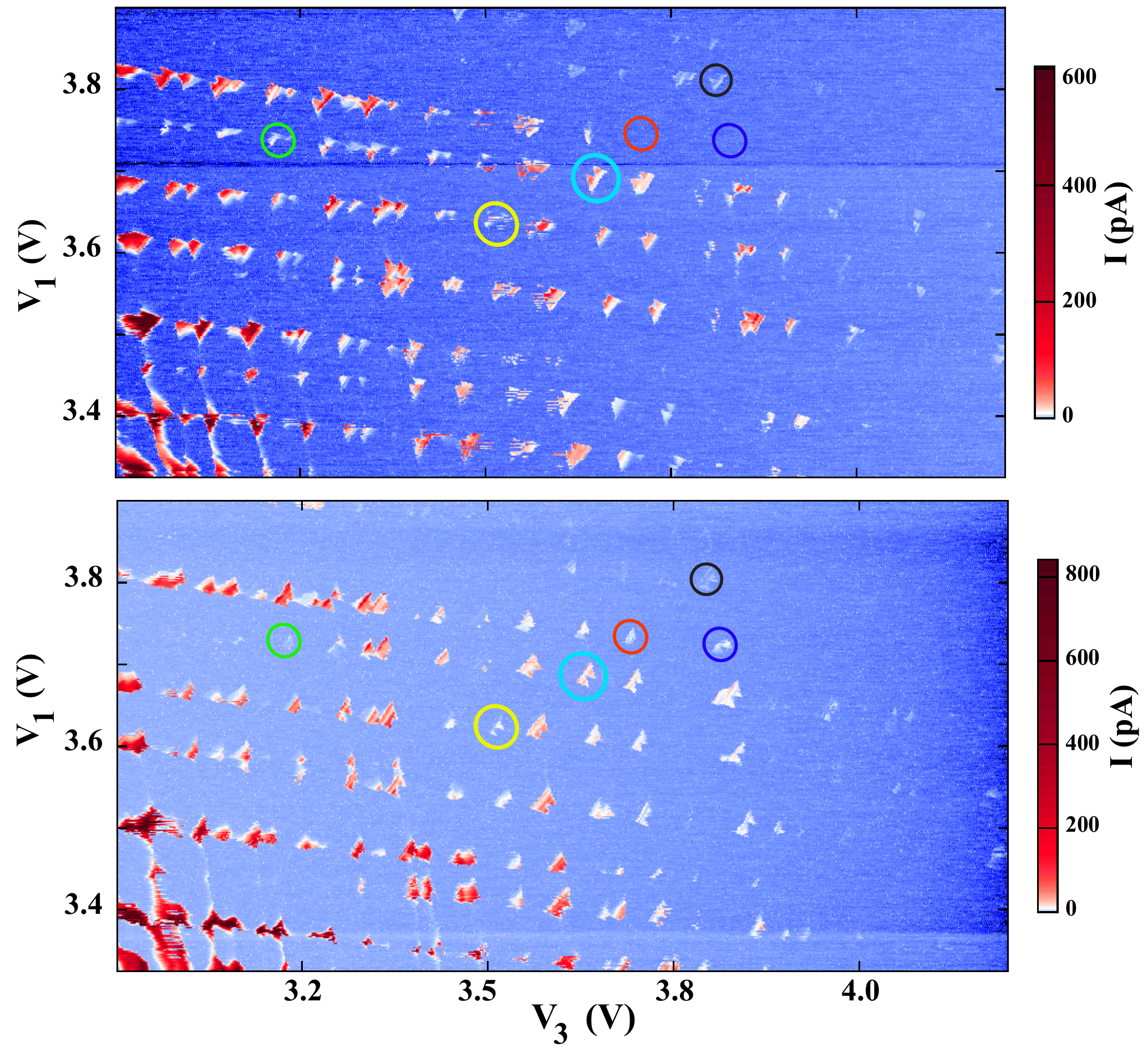}
\caption{Double quantum dot charge stability diagrams in opposite bias directions (top: $V_{SD}=4$~mV, bottom: $V_{SD}=-4$~mV). The plots show the absolute value of measured leakage current across the dots from one reservoir to the other while scanning G1 versus G3 at a fixed G2. Some charge transitions that show bias asymmetric behavior are circled.
\label{figS1}}
\end{figure}

\begin{figure}
\includegraphics[width=7cm]{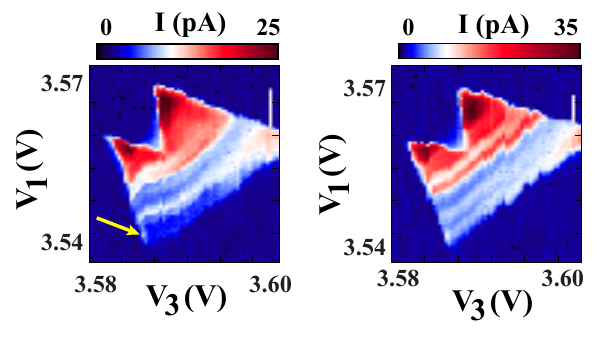}
\caption{Bias triangle of Fig.~3b in the main text at zero magnetic field (left) and at $B=-0.8$~T (right) at $V_{SD}=4$~mV. The current is suppressed at the base of the triangle (lower left) at zero field and there is an excess current at the side which can be representative of the hole exchange between the dot and the reservoir where their Fermi energy levels are equal. At finite magnetic field we can see the increase in the leakage current at the base of the triangle associated with lifting the spin blockade.
\label{figS1.1}}
\end{figure}

\begin{figure}[t]
\includegraphics[width=8cm]{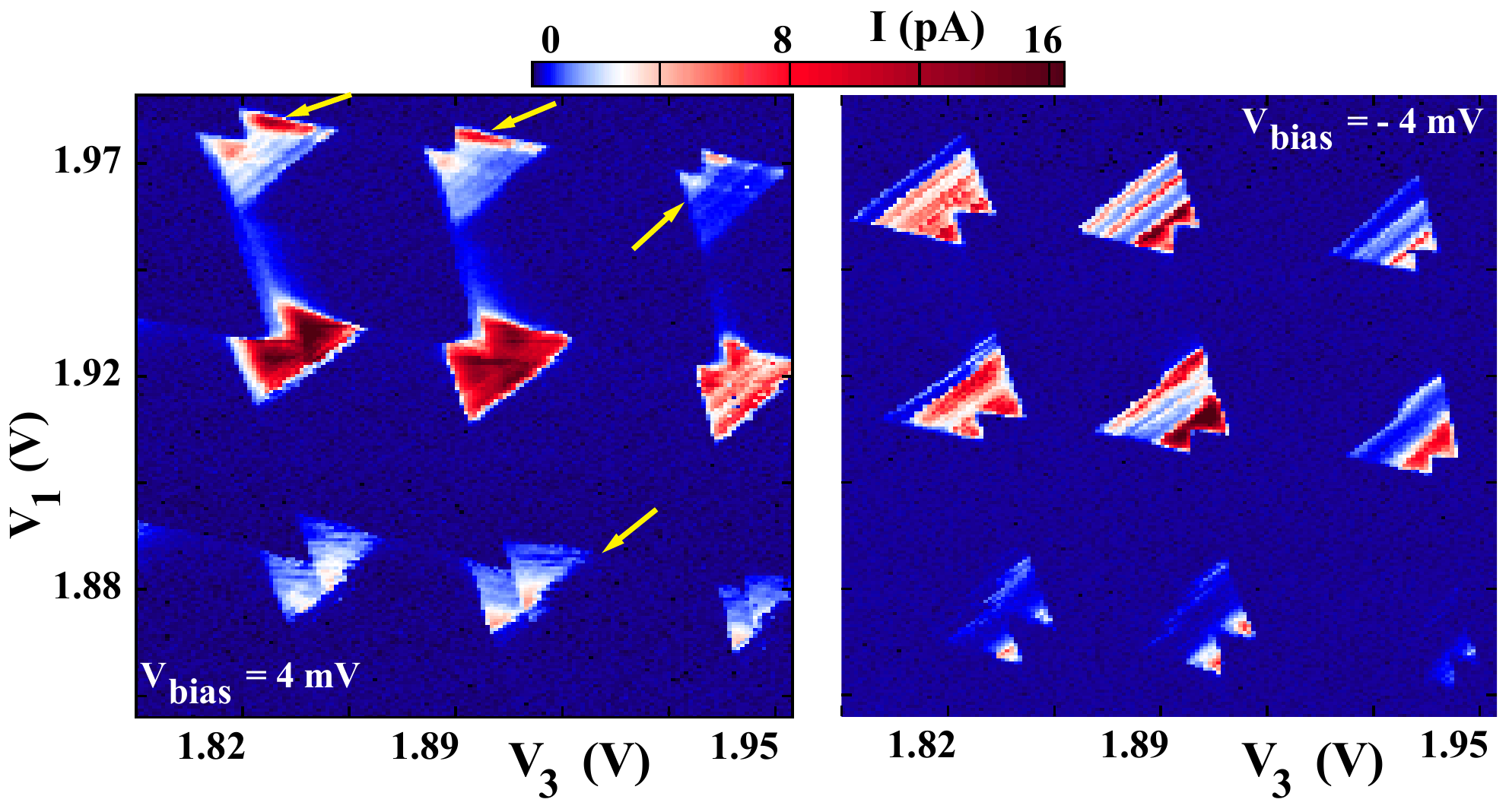}
\caption{Double quantum dot charge stability diagrams of Device B in opposite bias directions. The plots show the absolute value of measured leakage current across the dots from one reservoir to the other while scanning G1 versus G3 at a fixed G2. The arrows show characteristics of spin blockade explained in the text.
\label{figS2}}
\end{figure}

In Fig.~\ref{figS1} we show part of the charge stability diagram shown in Fig.~1 of the main text, in both source-drain bias directions and circle the charge transitions that could be candidates for Pauli spin blockade. Fig.~\ref{figS1.1} shows the bias triangle of Fig.~3b of the main text at zero and finite magnetic field, where the spin blockade is lifted up.
In Fig.~\ref{figS2} we see a few charge transitions from Device B in opposite bias directions. These transitions manifest some characteristics of Pauli spin blockade at zero magnetic field such as suppressed current at the base of the triangles, enhanced current on the side of the triangles related to spin exchange with the reservoir leads, and triplet hats corresponding to spin transitions of triplet states when they become energetically accessible.
However, these features by themselves cannot be conclusive and one needs to apply external magnetic field and study their field-dependent behavior, such as shown for Device A in the main text.

\begin{figure}
\includegraphics[width=5.5cm]{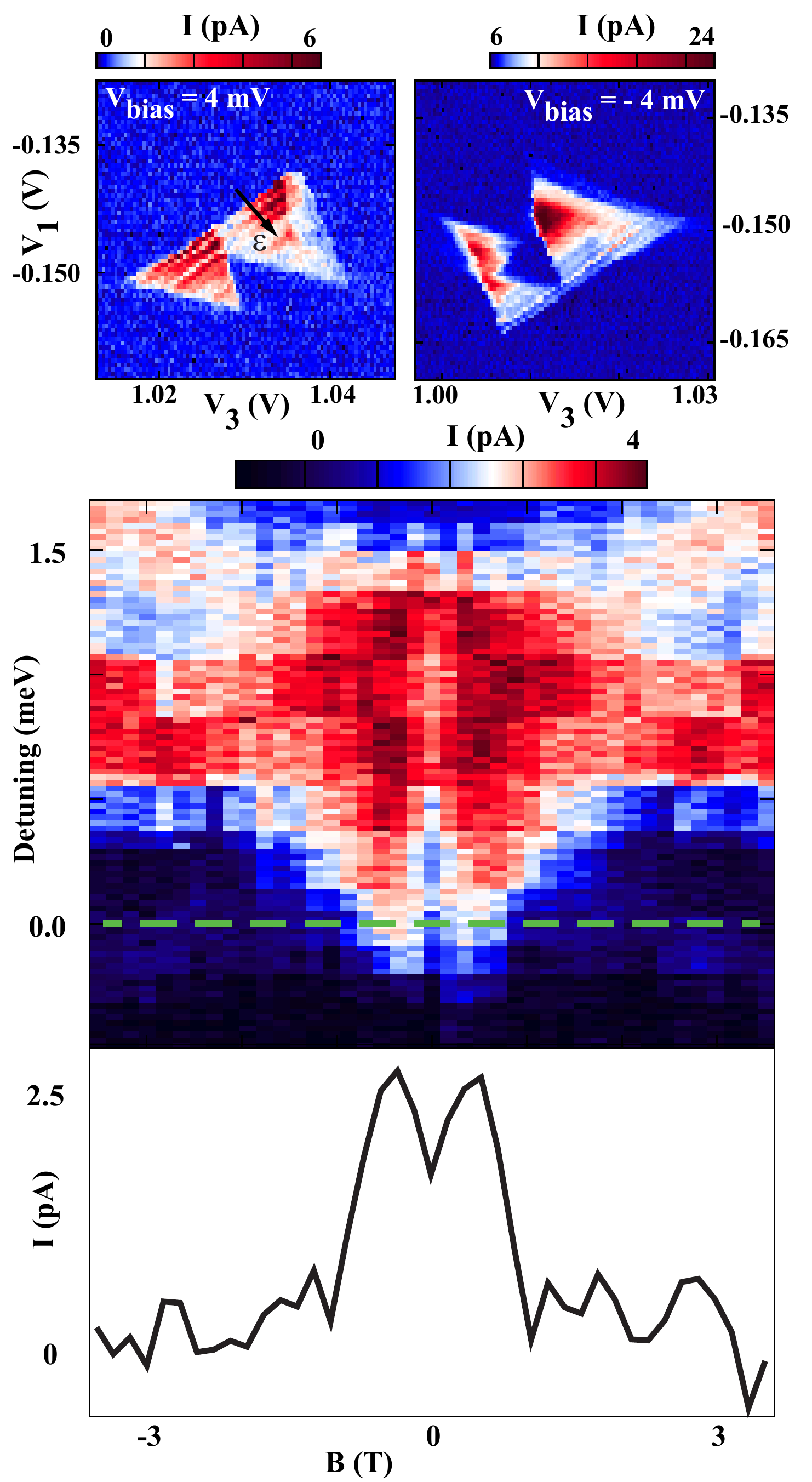}
\caption{Pauli spin blockade in Device C. Top: the bias triangle in two opposite source-drain bias directions. The triangle in the left is smaller than the one on the right showing that the current at its based is suppressed. The color plot is the absolute value of the leakage current. Bottom: leakage current through the double quantum dot measured as a function of detuning and magnetic field at $V_{SD}=4$~mV. The double-peak behavior highlighted by a line cut at zero detuning corresponds to spin blockade.
\label{figS3}}
\end{figure}

We can see the evolution of spin states in the the presence of a magnetic field from Device C shown in Fig.~\ref{figS3}. In the upper part of Fig.~\ref{figS3} we see the suppressed current at the base of the bias triangle at $V_{SD}=4$~mV. The double-peak structure is evident in the scan of magnetic field versus detuning, where spin blockade is lifted at finite fields.

\clearpage

\section{$g$-factor anisotropy}

\begin{figure}[h]
\includegraphics[width=11cm]{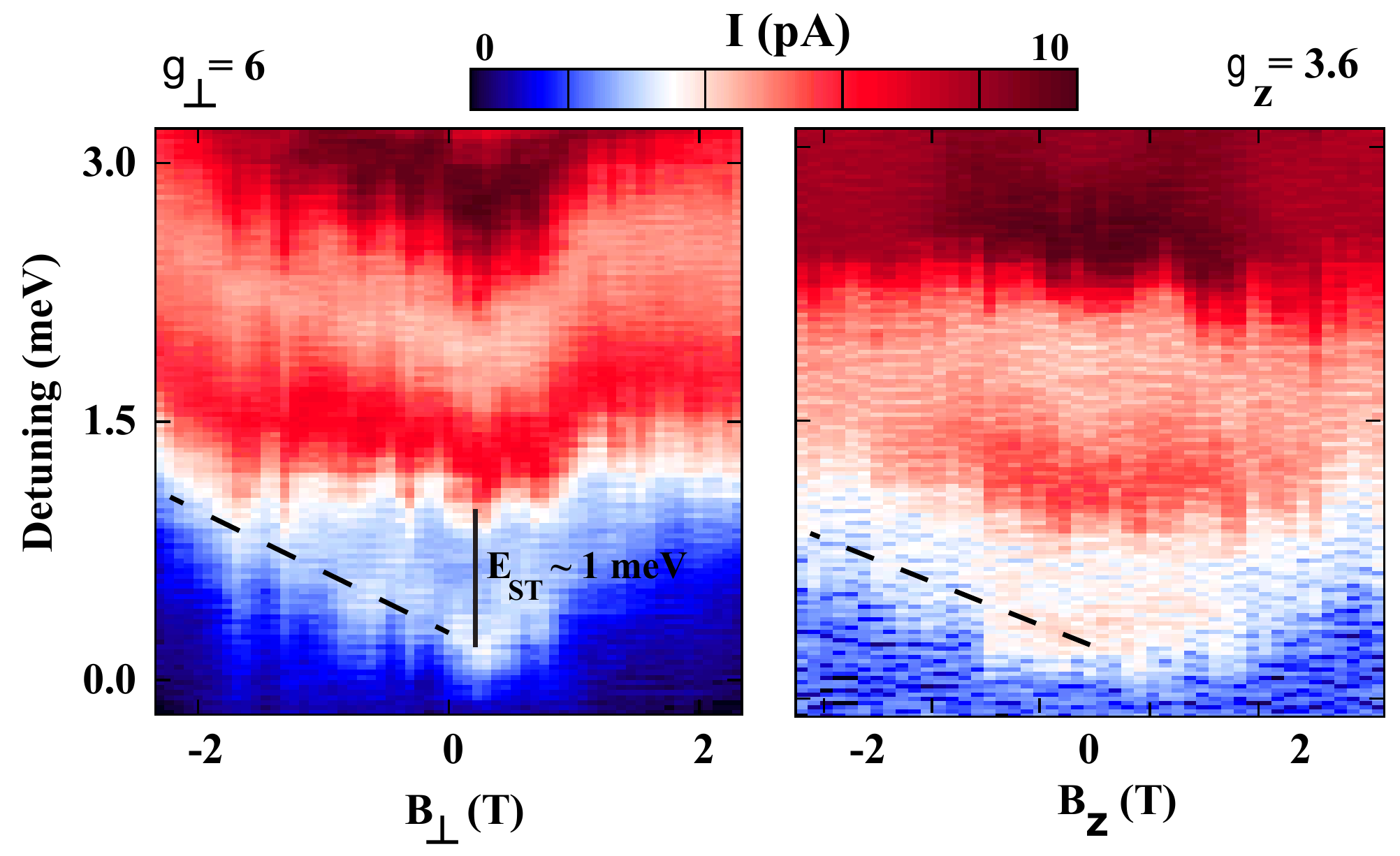}
\caption{Leakage current as a function of magnetic field and detuning. Left: field is applied normal to the substrate. Right: in-plane magnetic field. The dashed lines show the slope of the resonance line moving as a function of filed and detuning from which we measure the effective $g$-factor. The solid line is used to read the singlet-triplet energy splitting. It is not trivial for the right panel to draw a solid line for this energy splitting.
\label{figS4}}
\end{figure}

\begin{figure}[t]
\includegraphics[width=12cm]{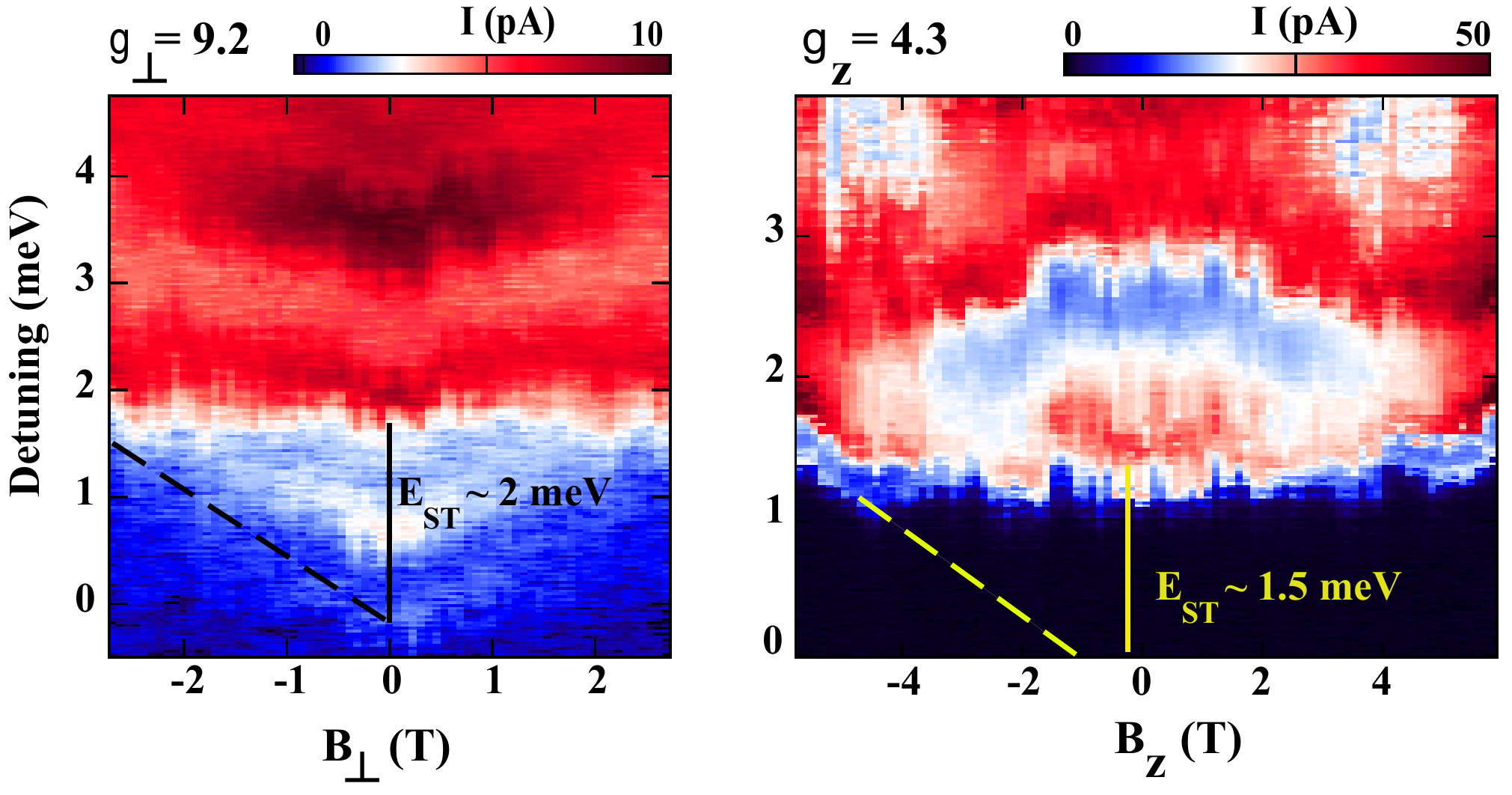}
\caption{Leakage current of the same triangle shown in Fig.~2 of the main text (left panel here) as a function of magnetic field and detuning, where the field is applied normal to and in-plane with the substrate for the left and right panels, respectively. Dashed lines are guides to the eyes to show the slope of the moving resonance lines in the present of the field. The current is extremely suppressed below $\varepsilon=1$~meV in the right panel, but the tail of the line can be seen just above $\varepsilon \approx 1$~meV and $B_Z = \pm 6$~T. The solid lines represent the singlet-triplet energy splitting.
\label{figS5}}
\end{figure}

\begin{figure}[h]
\includegraphics[width=12cm]{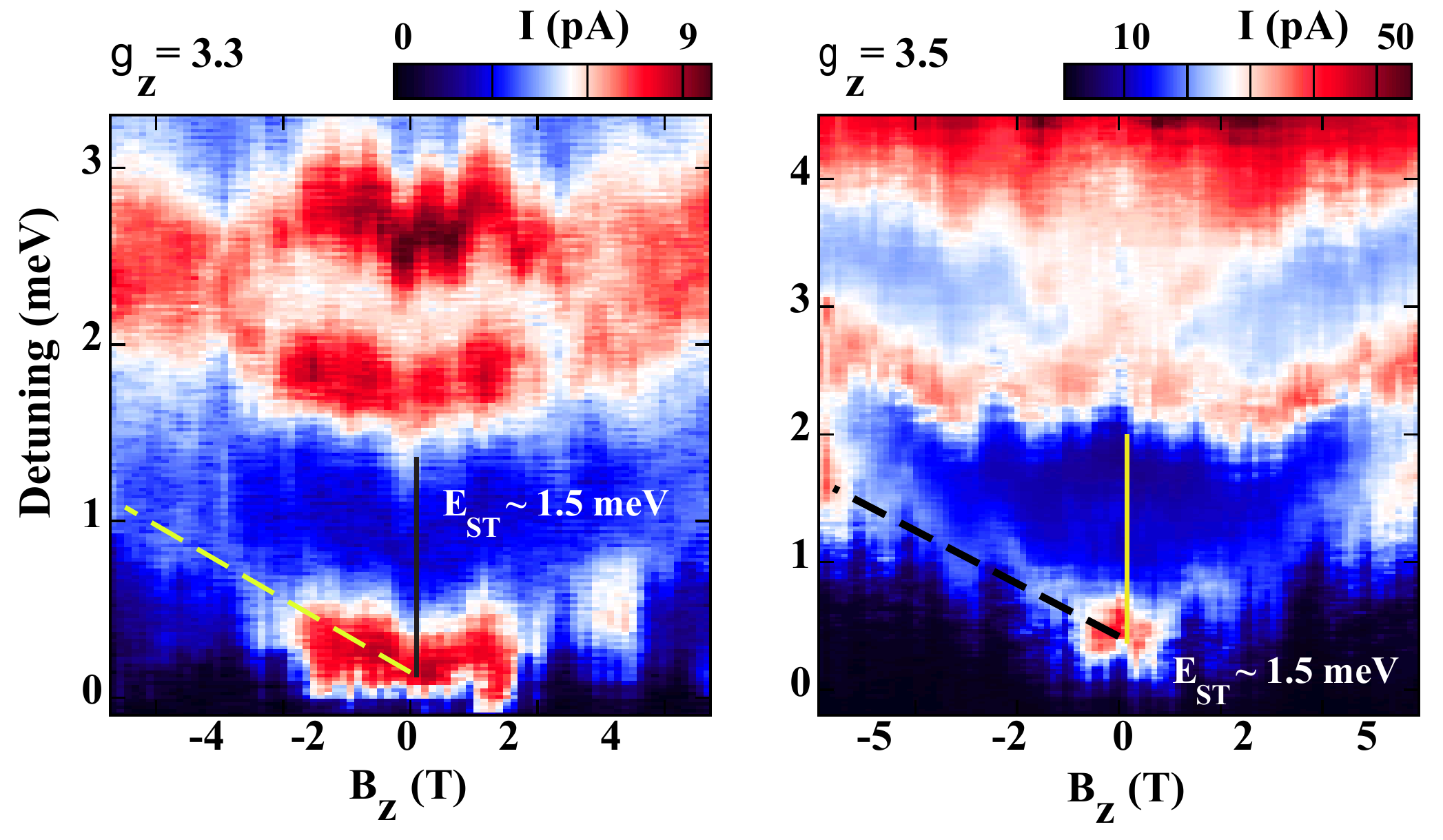}
\caption{Leakage current through two different double dot configurations both as a function of in-plane magnetic field and detuning. Dashed lines are used to measure the effective $g$-factors along the direction of the magnetic field, and the singlet-triplet energy splittings are shown by solid lines.
\label{figS6}}
\end{figure}

Below we show multiple scans of leakage current through different double quantum dot configurations as a function of magnetic field and detuning for Device A. As we see not all the scans reveal a sharp resonance line, yet an effective $g$-factor can be read off from the slope of the resonance associated with $T_+(1,1) \to S(0,2)$ transition, shown by dashed lines in Fig.~\ref{figS4}--Fig.~\ref{figS6}.

We apply magnetic fields in two different directions: (i) normal to the plane of the nanowire and local gates ($B_{\perp}$) and (ii) in-plane with the nanowire and gates, where the nanowire makes an angle of $\sim$~30$^{\circ}$ with the field ($B_z$). 

Fig.~\ref{figS4} and Fig.~\ref{figS5} show the leakage current through the double quantum dot in both directions of applied magnetic field. The slopes from which we can read the effective $g$-factors are different for the two cases of applied field, larger for the out of plane magnetic field and smaller for the in-plane field. This confirms the anisotropy of the $g$-factor, where $g_{\perp} \approx 7.6$ and $g_z \approx 4.03$.

Also from these scans we measure a singlet-triplet energy splitting $E_{ST} \sim 1$--$2$~meV.

\clearpage

\section{Full theory model, including a $g$-factor gradient}

The full model we used to produce the theoretical curves shown in Fig.~3 of the main text is based on the five-level Hamiltonian. In this section we set again $\hbar = k_{\rm B} = g\mu_{\rm B} = e = 1$.
\begin{align}
H = \left( \begin{array}{ccccc}
0 & iB & 0 & 0 & i\alpha t \\
-iB & 0 & 0 & 0 & i\alpha t \\
0 & 0 & 0 & \xi B & i\alpha t \\
0 & 0 & \xi B & 0 & t \\
-i\alpha t & -i\alpha t & -i\alpha t & t & 0
\end{array}\right),\label{eq:ham}
\end{align}
again written in the basis $\{ \ket{T_x}, \ket{T_y}, \ket{T_z}, \ket{S}, \ket{S_{02}} \}$, where $\ket{T_z} = \ket{T_0}$ and $\ket{T_{x,y}} = i^{1/2 \mp 1/2} \{\ket{T_-} \mp \ket{T_+}\}/\sqrt 2$.
The difference in Zeeman splittings in the two dots, caused by a finite $g$-factor gradient $\xi = \frac{1}{2}(g_L - g_R)/(g_L+g_R)$, results in a coherent mixing of $\ket{T_z}$ and $\ket{S}$, proportional to $\xi B$.

We diagonalize this Hamiltonian, denoting the resulting five eigenstates as $\ket{n}$, with $n=0\dots 4$ increasing with increasing eigenenergy.
In this basis we write a time-evolution equation for the density matrix,
\begin{align}
\frac{d\hat\rho}{dt} = -i[H^{\rm diag},\hat\rho ] + \boldsymbol \Gamma \hat\rho + \boldsymbol \Gamma_{\rm rel} \hat\rho.
\label{eq:drdt}
\end{align}
The coupling to the source and drain leads is described by the operator $\boldsymbol\Gamma$, including (i) the decay of all states $\ket{n}$ to the drain lead with the rates $\Gamma_n = \Gamma |\langle {n|S_{02}} \rangle|^2$ and (ii) immediate reload into one of the (1,1) states with equal probabilities.
Explicitly, this operator reads
\begin{align}
\big( \boldsymbol \Gamma \hat\rho \big)_{ij} =
-\frac{\Gamma_i+\Gamma_j}{2}\rho_{ij}
+\frac{1}{4} \left( \sum_n \Gamma_n \rho_{nn} \right) \left\{ 1 - |\langle {i|S_{02}} \rangle|^2 \right\} \delta_{i,j}.
\end{align}
The relaxation operator $\boldsymbol\Gamma_{\rm rel}$ describes dissipative relaxation in the simplest way possible: 
We assume that all four excited states relax to the ground state $\ket{0}$ with the same rate $\Gamma_{{\rm rel},n} = \Gamma_{\rm rel}$, and for the ground state $\Gamma_{{\rm rel},0} = 0$.
With this notation, we write the relaxation operator as
\begin{align}
\big( \boldsymbol \Gamma_{\rm rel} \hat\rho \big)_{ij} =
-\frac{\Gamma_{{\rm rel},i}+\Gamma_{{\rm rel},j}}{2}\rho_{ij}
+ \left( \sum_n \Gamma_{{\rm rel},n} \rho_{nn} \right) \delta_{i,0}\delta_{j,0}.
\end{align}
The ground state at $B=0$ is an equal superposition of $\ket{S_{02}}$ and the optimally coupled $(1,1)$ state $\ket{m} = \{\ket{S} - i\alpha \mathbbm{1}\cdot\ket{\vec T}\}/\sqrt{1+3\alpha^2}$, and for increasing $B$ it develops into a pure $\ket{T_+}$-state.
This means that the ground state always has a $(1,1)$-probability of at least 1/2.
The relaxation model used can thus be seen as a crude approximation for dissipative spin relaxation within the $(1,1)$ subspace.

We can then solve Eq.~\ref{eq:drdt} in steady state, $d\hat\rho / dt = 0$, and find the current from the resulting equilibrium occupation probabilities $p_n = \hat\rho_{nn}$ as $I = \sum_n p_n\Gamma |\langle {n|S_{02}} \rangle|^2$.
The result for $\xi = 0$ in the limit of $\gamma \ll 1$ is given in the main text; for finite $\xi$ we arrive at the more complicated expression
\begin{align}
I(B) = \frac{4 \Gamma}{\Omega}\Big\{ {} & {}
t^2 \left[B^2 (1+\gamma) \xi ^2+t^2\gamma  \zeta_+ \right] \left[B^2 \zeta_- (1-\xi ^2)-\zeta_+ (\tau ^2+w)\right] 
\nonumber\\ {} & {} 
\times \left[t^2 \zeta_+ \left(\tau ^2-w(1+4 \gamma) \right)-B^2 \left( t^2\zeta_-(1-\xi ^2) +4 w \gamma  \xi ^2 \right)\right]
\Big\},
\label{eq:curr}
\end{align}
with
\begin{align}
\Omega = {} & {} 4 B^6 (1+\gamma) \left[t^4\zeta_-^2(1-\xi ^2)^2 \xi ^2  +16w^2 \gamma  \xi ^6 \right]
\nonumber\\ {} & {}
+B^4 t^2 \zeta_+ \big[16w^2 (1+2 \gamma) (1+6 \gamma) \xi ^4 - t^2\zeta_- (1-\xi ^2)
\nonumber\\ {} & {}
\hspace{5em}\times\left(\tau ^2 (7+ 12 \gamma ) \xi ^2 + t^2 (1-4 \gamma )(1+\alpha^2+2 \alpha ^2 \xi ^4) \right)\big]
\nonumber\\ {} & {}
+2 B^2 t^4 \zeta_+^2 \big[\tau ^4 (1 + 6 \gamma) \xi ^2 +t^2 \tau ^2(1-4 \gamma )  (1+\alpha^2+ 2 \alpha ^2 \xi ^4)
\nonumber\\ {} & {}
\hspace{5em}+ 6 w^2 (1+ 7\gamma +16 \gamma^2) \xi ^2\big]
\nonumber\\ {} & {}
+ t^6 \zeta_+^3\big[  w^2 (1+ 4 \gamma) (1+ 16 \gamma)-\tau ^4(1-4 \gamma ) \big],
\label{eq:omega}
\end{align}
where we use again the notation $w = \sqrt{(B^2 - \tau^2)^2 + 8B^2\alpha^2t^2}$, $\gamma = \Gamma_{\rm rel} / \Gamma$, and $\tau = t \sqrt{1 + 3\alpha^2}$;
in addition we defined $\zeta_\pm = 1+\alpha^2(1\pm 2\xi^2)$.
We again emphasize that, in contrast with Eq.~3 in the main text, while deriving this expression we did \emph{not} assume $\gamma \ll 1$.
In the limit of $\xi \to 0$ and $\gamma \ll 1$ Eq.~\ref{eq:curr} reduces indeed to Eq.~3 of the main text.

\section{Theory curves in Fig.~3}

Using Eq.~3 from the main text (where $\xi = 0$), we can already model the zero-detuning current traces shown in the right part of Fig.~3 of the main text.
When we attempt to fit the model to the data, we find that in all cases the overall magnitude of the current can be related straightforwardly to $\Gamma_{\rm rel}$, but the other model parameters ($t$, $\alpha$, and $\Gamma$) cannot be determined independently.
There are too many model parameters as compared to the number of ``distinguishing features'' in the experimental current traces, i.e., one needs to fix at least one of the remaining parameters to get anywhere close to a reasonable fit:
For instance, we can vary $\Gamma$ over several orders of magnitude, and for each value we can produce a reasonable fit fixing $\tau$ and $\alpha$.

Although more complicated and containing even one more parameter, the model with finite $\xi$ is in a way easier to fit to the data.
Given that we can safely assume that $\xi \approx 2$--5\% ($g$-factors are known to vary from dot to dot in Ge/Si nanowires, and we have a rough upper bound on $\xi$ from Fig.~2 in the main text), we find that the resulting finite Zeeman gradient over the dots plays an important role:
Since it mixes $\ket{T_z}$ and $\ket{S}$, it provides a way out of the single blocked state left at finite field $\{\ket{T_z} + i\alpha \ket{S}\}/\sqrt{1+\alpha^2}$.
The associated $\xi$-induced decay rate $\Gamma_{\xi} \sim (\xi B)^2\Gamma / t^2$ competes with $\Gamma_{\rm rel}$, and when $\Gamma_\xi \gtrsim \Gamma_{\rm rel}$ then the overall scale of the current will be set by $\Gamma_\xi$ instead of $\Gamma_{\rm rel}$.
The typical magnitude of the current (several pA for the data) thus sets an extra constraint on $\Gamma$.
With a $\xi$ on the order of a few percent, this provides information about $\Gamma$ and helps the fitting procedure to converge.

\begin{figure}[t]
\begin{center}
\includegraphics[scale=1]{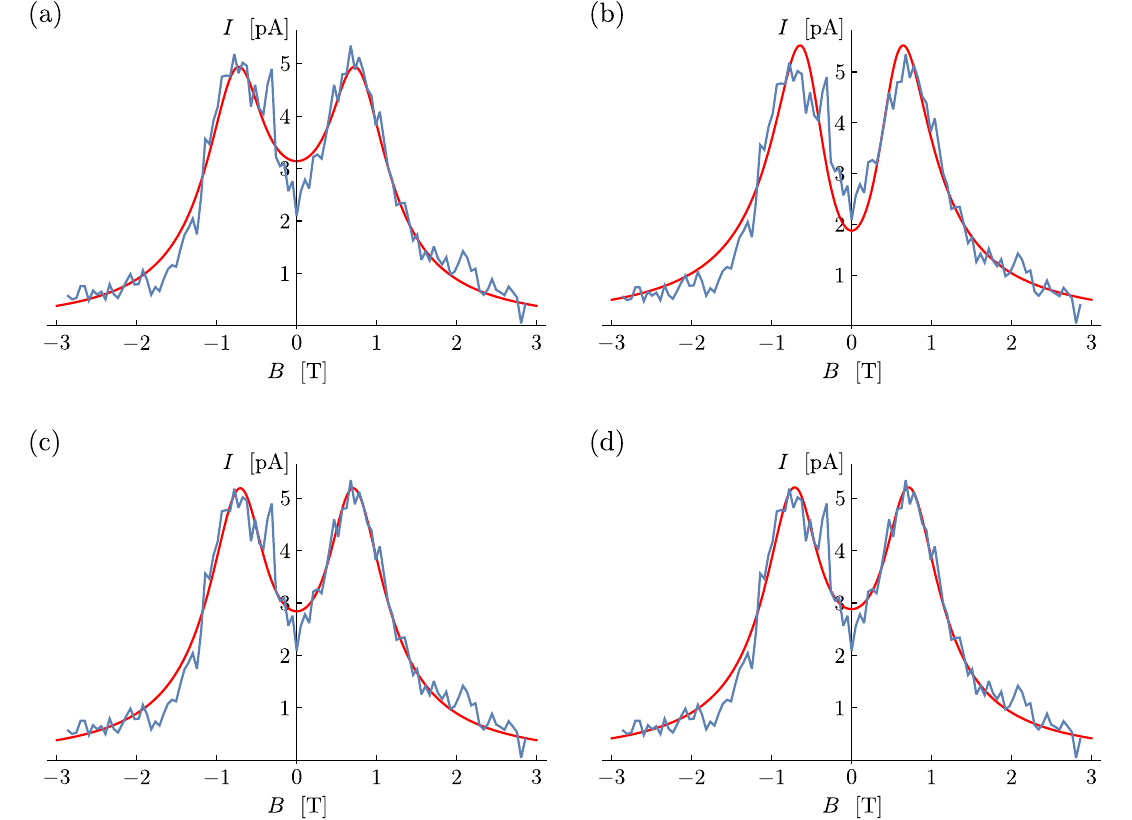}
\caption{Series of fits (red curves) of the double-peak data of Fig.~3b of the main text (blue curves) to Eqs.~\ref{eq:curr}--\ref{eq:omega}, illustrating the freedom in parameter space.
(a,b) Least-squares fit with (a) $\xi = 0.03$ and $\Gamma = 200$~MHz and (b) $\xi = 0.03$ and $\Gamma = 1$~GHz.
(c,d) Least-squares fit with (c) $\xi = 0.02$ and (d) $\xi = 0.05$.
All resulting fit parameters are detailed in Table~\ref{tab:1}.}\label{fig:sm_jd_1}
\end{center}
\end{figure}

\begin{table}[t]
\begin{center}
\begin{tabular}{lccccc}
\hline
Fit & $\xi$ & $\Gamma$~[MHz] & $\alpha$ & $t~[\mu$eV] & $\gamma$ \\
\hline
Fig.~3b & \textit{0.03} & $256 \pm 19$ & $0.37 \pm 0.02$ & $165 \pm 5$ & $0.061 \pm 0.007$ \\
Fig.~S7a & \textit{0.03} & \textit{200} & $0.41 \pm 0.03$ & $169 \pm 6$ & $0.092 \pm 0.002$ \\
Fig.~S7b & \textit{0.03} & \textit{1000} & $0.20 \pm 0.01$ & $154 \pm 4$ & $0.0090 \pm 0.0002$ \\
Fig.~S7c & \textit{0.02} & $261 \pm 20$ & $0.37 \pm 0.02$ & $165 \pm 5$ & $0.059 \pm 0.007$ \\
Fig.~S7d & \textit{0.05} & $244 \pm 16$ & $0.36 \pm 0.02$ & $166 \pm 5$ & $0.065 \pm 0.007$ \\
\hline
\end{tabular}
\end{center}
\caption{Fit parameters found for all fits to the double-peak data shown in the main text (first row) and the supplementary (other rows). In all cases Eqs.~\ref{eq:curr}--\ref{eq:omega} presented above were used for fitting. Italic numbers indicate values that were fixed before fitting.}\label{tab:1}
\end{table}

We first focus on the double-peak data of Fig.~3b and try to fit the model of Eqs.~\ref{eq:curr}--\ref{eq:omega} to the data.
Fixing $\xi = 0.03$, we can obtain reasonable fits for values for $\Gamma$ ranging from 0.2--1~GHz.
The best fit (i.e., a least-squares fit) is the one we plotted in Fig.~3b, yielding $\Gamma = 256$~MHz, $t = 165~\mu$eV, $\gamma = 0.061$, and $\alpha = 0.37$ (where we used the average $g = 4.4$, determined from the slope of the base line of the bias triangle).
Over the whole range of $\Gamma = 0.2$--1~GHz, we find $t = 169$--$154~\mu$eV, $\gamma = 0.092$--0.009, and $\alpha = 0.41$--0.20; varying $\xi$ within the range $0.02$--$0.05$ does not make a large difference for the fit parameters we find.
In Fig.~\ref{fig:sm_jd_1} we present several different fits covering the mentioned range of parameters: Fig.~\ref{fig:sm_jd_1}a shows the best fit obtainable for $\Gamma = 200$~MHz and Fig.~\ref{fig:sm_jd_1}b that for $\Gamma = 1$~GHz (a complete overview of all detailed fit parameters is given in Table~\ref{tab:1}.
In Figs.~\ref{fig:sm_jd_1}c,d we show the best fits we could produce using $\xi=0.02$ and $\xi = 0.05$ respectively, see again Table~\ref{tab:1} for all fit parameters.
Based on these results, we conclude that most likely $\alpha$ is of the order $\sim 0.3$ for this data set, which would signal a significant effect of (pseudo-)spin-orbit coupling on the low-energy dynamics of this double-dot system.

The more noisy and less feature-rich single-peak data of Fig.~3a are still hard to fit:
Even including the $g$-factor gradient we cannot reasonably narrow down all fit parameters to a range of values.
In Fig.~3a we plotted two different theory curves for $\xi = 0.03$ on top of the data.
For both curves we added a constant current of $0.8~$pA to account for the background signal observed in the data, and we have fixed $\alpha = 0.4$, which is inside the range of $\alpha$'s extracted from the double-peak data.
For the solid red curve we have set $\Gamma = 300$~MHz (not far from the same value as found above) and then tuned $t = 55~\mu$eV and $\gamma = 0.007$ to match the data best (with $g = 5$, again determined from the slope of the base line).
The double-peak structure that is clearly visible in the theoretical curve cannot be seen in the data, and is on the edge of being too pronounced to agree qualitatively with the data.
The appearance of the theory curve can improve if we decrease $\Gamma$, as shown by the green dashed curve where $\Gamma = 25$~MHz, $t = 170~\mu$eV, and $\gamma = 0.7$.
Although most of the parameters now seem to have entered relatively unlikely regimes, we emphasize again that the detailed (spin) physics underlying the hole Pauli blockade in Ge/Si nanowires are not well understood yet, and our intuition for all model parameters is mostly based on the spin physics of conduction electrons in III-V semiconductors.

We finally point out that the fact that our fitting procedure in this case does not straightforwardly yield a single-peak curve such as the one in Fig.~4 of the main text, is caused by the $g$-factor gradient we added to the model.
A finite $\xi$ puts an upper bound on $\Gamma$ in our model based on the magnitude of the current.
This helps the fit of the double-peak data converge, but also keeps us away from the single-peak regime of the model with $\xi=0$:
For the zero-field dip to become indistinguishably small, a large parameter $\beta \sim \alpha \sqrt{\Gamma/\Gamma_{\rm rel}}$ is required;
the current level determines $\Gamma_{\rm rel}$, which, together with the bound on $\Gamma$, prevents us from reaching the large-$\beta$ regime.
\end{document}